\documentclass[12pt,preprint]{aastex}
\usepackage{natbib}
\usepackage{graphicx}
\shorttitle{Constraining annihilating dark matter}
\begin{document}
\title{Constraining annihilating dark matter using the multi-frequency radio flux profiles of the M33 galaxy}
\author{$^a$Man Ho Chan\thanks{chanmh@eduhk.hk} , $^a$Chak Man Lee, $^b$Lang Cui\thanks{cuilang@xao.ac.cn} , $^b$Ning Chang and $^c$Chun Sing Leung}
\affil{$^a$ Department of Science and Environmental Studies, The Education University of Hong Kong, Tai Po, Hong Kong, China
\\ $^b$ Xinjiang Astronomical Observatory, Chinese Academy of Sciences, Urumqi, China
\\ $^c$ Department of Applied Mathematics, Hong Kong Polytechnic University, Hong Kong, China}

\begin{abstract}
Radio data can give stringent constraints for annihilating dark matter. In general, radio observations can detect very accurate radio flux density with high resolution and different frequencies for nearby galaxies. We are able to obtain the radio flux density as a function of distance from the galactic center and frequencies $S(r,\nu)$. In this article, we demonstrate a comprehensive radio analysis of the M33 galaxy, combining the radio flux density profile $S(r)$ and the frequency spectrum $S(\nu)$ to get the constraints of dark matter annihilation parameters. By analyzing the archival radio data obtained from the Effelsberg telescope, we show that the dark matter annihilation contributing to the radio flux density might be insignificant in the disk region of the M33 galaxy. Moreover, by including the baryonic radio contribution, we constrain the $2\sigma$ conservative upper limits of the annihilation cross section, which can be complementary to the existing constraints based on neutrino, cosmic-ray, and gamma-ray observations. Our results indicate that analyzing the galactic multi-frequency radio flux profiles can give useful and authentic constraints on dark matter for the leptophilic annihilation channels.
\end{abstract}


\section{Introduction}
Recent studies using radio data of galaxies and galaxy clusters can constrain the dark matter annihilation parameters, such as the lower limits of dark matter mass and the upper limits of annihilation cross section for different annihilation channels \citep{Colafrancesco,Egorov,Storm,Chan,Chan2,Regis,Beck,Lavis}. If dark matter particles can self-annihilate, they could give high-energy electrons, positrons, photons, and neutrinos. In particular, the high-energy electrons and positrons can produce synchrotron radiation in radio bands when they are moving inside the magnetic field of a galaxy or a galaxy cluster. The underlying physics is well-known and this indirect method of dark matter detection has been applied for almost two decades \citep{Colafrancesco,Chan2,Beck,Lavis}.

Comparing with using gamma-ray data to constrain dark matter \citep{Calore,Daylan}, radio detection can provide very accurate radio flux density maps which indicate the radio flux density as a function of radius $r$ (i.e. the distance from galactic center) and frequencies $\nu$. Since the resolution of gamma-ray detection is relatively poor, it is very difficult to obtain the gamma-ray flux as a function of $r$ for other galaxies or galaxy clusters. For radio analyses, especially in galaxies, some previous studies have used the radio spectrum $S(\nu)$ \citep{Colafrancesco,Chan3,Chan4} or the radio flux density profile $S(r)$ \citep{Chan5} to constrain dark matter parameters. These methods can provide different perspectives in handling the radio contribution of dark matter annihilation in galaxies.

In this article, by using the archival radio data of the M33 galaxy \citep{Buczilowski2,Buczilowski,Tabatabaei}, we re-construct the radio maps of the M33 galaxy for different observing frequencies. Then, we can obtain the radio flux density as a function of both radius $r$ and frequency $\nu$: $S(r,\nu)$. Based on this multi-variable function, we have developed a comprehensive radio analysis by combining the frequency spectral data (flux with different frequencies) and radio flux density profiles (flux at different $r$) to constrain the dark matter parameters. Some recent studies have applied this similar idea to analyze the multi-frequency radio flux density profiles to constrain annihilating dark matter in the Large Magellanic Cloud \citep{Regis} and galaxy clusters \citep{Beck}. This study provides the first analysis of the M33 galaxy by using this comprehensive method.

\section{Radio flux profile construction}
We have used the archival radio data of the M33 galaxy obtained by the Effelsberg radio telescope in \citet{Buczilowski2,Buczilowski,Tabatabaei}. The data consist of four different central frequencies (1.42 GHz, 2.70 GHz, 4.85 GHz and 8.35 GHz) at different positions of the M33 galaxy. After performing the data reduction, we can plot the radio flux density maps (in mJy/beam) for different frequencies (see Fig.~1 for the map with $\nu=4.85$ GHz).

However, the radio flux density maps might consist of some strong foreground or background point sources which should not be included. Therefore, we subtract the bright compact point sources from the radio flux density maps to get the reduced radio flux maps. Here, since we do not have enough point source information for the 1.42 GHz radio flux density map, we cannot get the reduced radio flux map for $\nu=1.42$ GHz. We will only analyse the other three radio flux density maps (2.70 GHz, 4.85 GHz, and 8.35 GHz) in the followings.

Using a constant bin of $r=60$ arcsec, we take the azimuthal averaging of the radio flux density in concentric bins for different frequencies from the reduced radio flux maps. The fluctuations in the radio flux density for the same bin would contribute to the $1\sigma$ error bar of the data. Following this method, we can get the radio flux density profiles $S(r)$ for $\nu=2.70$ GHz, 4.85 GHz and 8.35 GHz (see Fig.~2). These three radio flux density profiles can be combined to form the multi-variable function $S(r,\nu)$.

M33 has a bulge component and a disk component. However, the radio flux density profiles only include one data point for the bulge component. To minimize the free parameters in our analysis, we will only analyze the disk region ($r>96"$), but not the data points for the bulge component. For the disk region, we have total 19 data points for each of the observing frequencies in the radio flux density profiles.

\begin{figure}
\begin{center}
\vskip 3mm
\includegraphics[width=140mm]{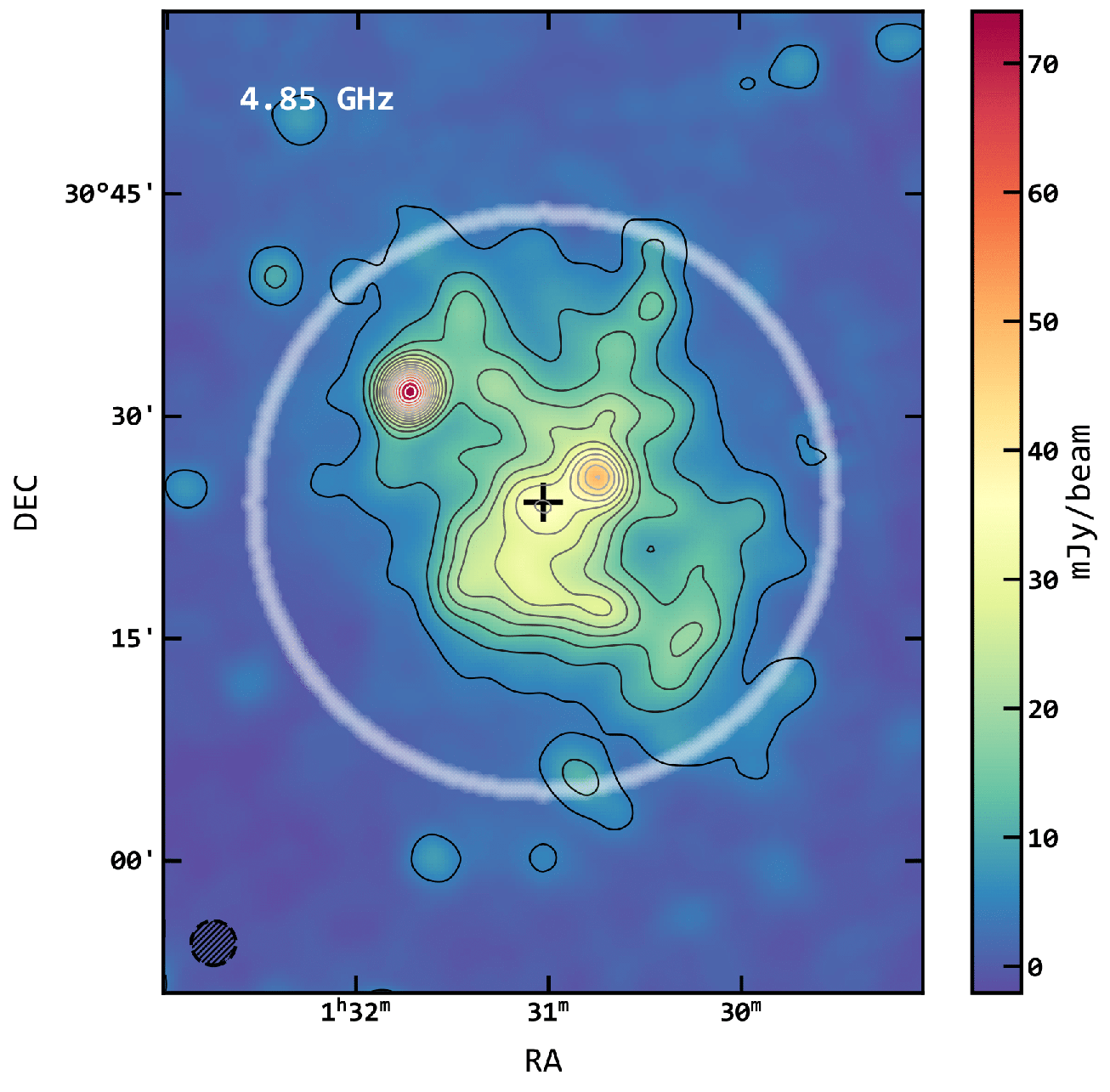}
\caption{The radio flux density map of the M33 galaxy with $\nu=4.85$ GHz. The region inside the white circle defines our interested region of the M33 galaxy. The colours in the map indicate different radio flux density in mJy/beam.}
\label{Fig1}
\end{center}
\end{figure}

\begin{figure}
\begin{center}
\vskip 3mm
\includegraphics[width=140mm]{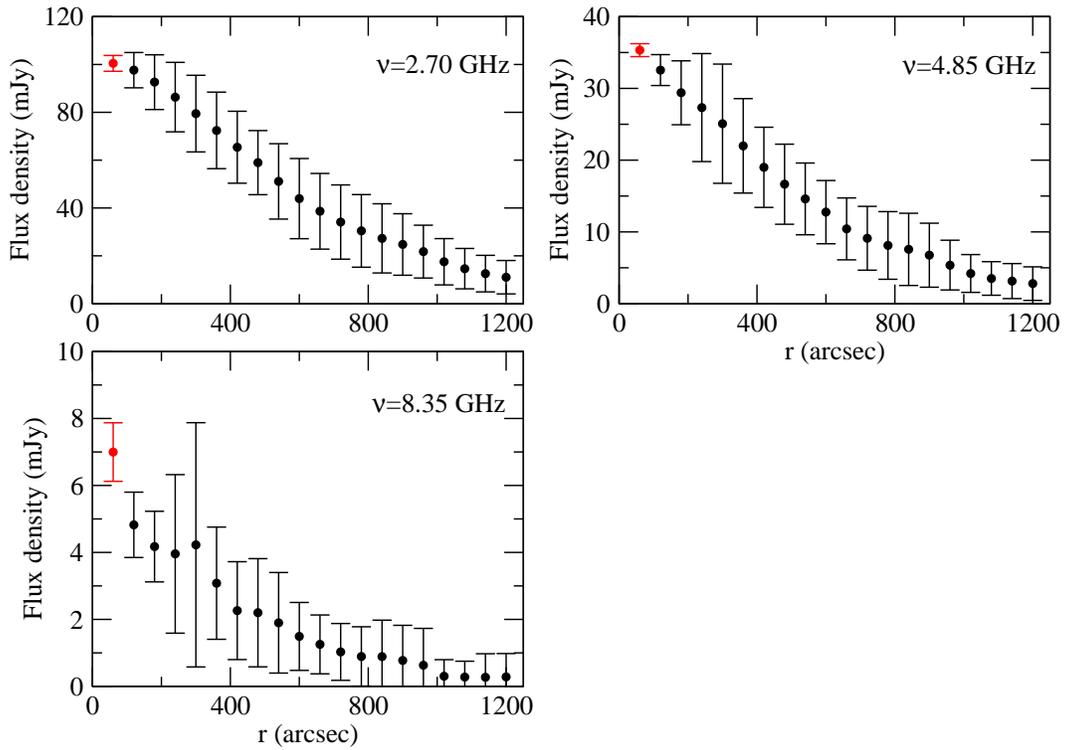}
\caption{The radio flux density (per beam size) profiles of the M33 galaxy for $\nu=2.70$ GHz, 4.85 GHz and 8.35 GHz. The red data points indicate the radio flux density in the bulge region while the black data points indicate the radio flux density in the disk region.}
\label{Fig2}
\end{center}
\end{figure}

\section{The theoretical framework}
A large amount of high-energy electrons and positrons would be produced via dark matter annihilation. The injection energy spectrum of these electrons and positrons $dN_{\rm e,inj}/dE$ for different annihilation channels can be predicted by numerical simulations \citep{Cirelli}. These high-energy electrons and positrons would emit synchrotron radiation in radio bands when there is a large magnetic field strength. The magnetic field strength in M33 is constrained to be $B=8.1 \pm 0.5$ $\mu$G \citep{Berkhuijsen}, which is relatively large among the galaxies in our Local Group. The cooling and the diffusion of the electrons and positrons produced from dark matter annihilation is given by the diffusion-cooling equation \citep{Ginzburg,Atoyan}:
\begin{eqnarray}
\frac{\partial}{\partial t}\frac{dn_{\rm e}}{dE}&=&\frac{D(E)}{r^2}\frac{\partial}{\partial r}\left(r^2\frac{\partial }{\partial r}\frac{dn_{\rm e}}{dE}\right)+\frac{\partial}{\partial E}\left[b_{\rm T}(E)\frac{dn_{\rm e}}{dE}\right]\nonumber\\
&&+Q(E,r),
\label{diffusion}
\end{eqnarray}
where $dn_e/dE$, $D(E)$ and $Q(E,r)$ are the electron/positron density spectrum, the diffusion function, and the particle-injection source, respectively. The diffusion function can be written as $D(E)=D_0(E/1\;{\rm GeV})^{\delta}$. We adopt the diffusion coefficient $D_0 = 2.0\times 10^{28} {\rm cm}^2{\rm s}^{-1}$ obtained in \citet{Berkhuijsen} and use the benchmark value of $\delta=1/3$ \citep{Kolmogorov} in our present study to model the diffusion process. In most galaxies, the cooling process is dominated by the synchrotron emission and inverse Compton scattering (ICS). The cooling function (in unit of $10^{-16}$ GeV s$^{-1}$) can be expressed as
\begin{equation}
b_{\rm T}(E)=0.0254E^2B^2 + U_{\rm rad}E^2.
\label{cooling}
\end{equation}
Here, $E$, $B$ are in the units of GeV and $\mu$G, respectively, and $U_{\rm rad}$ is the energy density of the Interstellar Radiation Field (ISRF) in the unit of ${\rm eV~cm}^{-3}$. For the ICS in M33, by adding the total radiation density ranging from infra-red to ultraviolet 0.77 eV cm$^{-3}$ \citep{Thirlwall} to the cosmic microwave background energy density 0.26 eV cm$^{-3}$, we get $U_{\rm rad} \approx 1.03$ eV cm$^{-3}$.

The particle-injection source term in Eq.~(\ref{diffusion}) is given by  \citep{Vollmann}
\begin{equation}
Q(E,r)=\frac{\langle \sigma v \rangle [\rho_{\rm DM}(r)]^2}{2m_{\rm DM}^2}\frac{dN_{\rm e,inj}}{dE},
\end{equation}
where $\rho_{\rm DM}(r)$ is the dark matter density in spherically symmetric profile and $\langle \sigma v \rangle$ is the annihilation cross section. By setting $\frac{\partial}{\partial t}(\frac{dn_{\rm e}}{dE})=0$ in Eq. (\ref{diffusion}) and satisfying the boundary condition $\frac{dn_{\rm e}(r_{\rm h},E)}{dE}=0$ with diffusion halo radius $r_{\rm h}$, the general solution of the equilibrium electron density spectrum is obtained in terms of the Fourier-series representation of the Green's function as follows \citep{Vollmann}:
\begin{eqnarray}
\frac{dn_{\rm e}}{dE}(E,r)&=&\sum_{n=1}^{\infty}\frac{2}{b_{\rm T}(E,r)r_{\rm h}}\frac{\sin\left(\frac{n\pi r}{r_{\rm h}}\right)}{r}
\nonumber\\
&&\times\int_E^{^{m_{\rm DM}}}dE'e^{-n^2[\eta(E)-\eta(E')]}\nonumber\\
&&\times\int_0^{r_{\rm h}}dr'r'\sin\left(\frac{n\pi r'}{r_{\rm h}}\right)Q(E',r'),
\label{dndE}
\end{eqnarray}
where the dimensionless variable $\eta(E)$ is given by
\begin{eqnarray}
\eta(E)&=&\frac{1}{1-\delta}\left(\frac{6.42\pi\;{\rm kpc}}{r_{\rm h}}\right)^2\left(\frac{D_0}{10^{28}{\rm cm^2/s}}\right)
\nonumber\\
&&\times\left(\frac{1}{1+\left({B}/{3.135 \;\mu{\rm G}}\right)^2}\right)\left(\frac{1\;{\rm GeV}}{E}\right)^{1-\delta}.
\end{eqnarray}
In our analysis, we take the halo size as the size of the M33 galaxy $r_{\rm h}=5.04$ kpc.

The average power at frequency $\nu$ under magnetic field $B$ for synchrotron emission induced by the dark matter annihilation is given by
\begin{equation}
P_{\rm syn}(\nu)=\int_0^{\pi}d\theta\frac{(\sin\theta)^2}{2}2\pi\sqrt{3}r_{\rm e}m_{\rm e}c\nu_{\rm g}F_{\rm syn}\left(\frac{x}{\sin\theta}\right),
\end{equation}
where $\nu_{\rm g}=eB/(2\pi m_{\rm e}c)$, $r_{\rm e}$ is the classical electron
radius, and $F_{\rm syn}(x/\sin\theta)=x/\sin\theta\int_{x/\sin\theta}^{\infty}K_{5/3}(y)dy$ with the quantity $x$ defined as
\begin{equation}
x=\frac{2\nu}{3\nu_{\rm g}\gamma^2}\left[1+\left(\frac{\gamma\nu_{\rm p}}{\nu}\right)^2\right]^{3/2},
\end{equation}
in which $\gamma$ is the Lorentz factor of the electron/positrons, and $\nu_{\rm p}=8890[n(r)/1\;{\rm cm}^{-3}]^{1/2}$ Hz is the plasma frequency with the number density of the thermal electrons ($n(r)\approx1\;{\rm cm^{-3}}$). The radio flux density emitted by a galaxy within a solid angle $\Delta \Omega$ due to dark matter annihilation as observed from Earth is:
\begin{equation}
S_{\rm DM}(\nu)=
2\frac{\Delta\Omega}{4\pi}\int ds\int_{m_{\rm e}}^{m_{\rm DM}}P_{\rm syn}(\nu)\frac{dn_{\rm e}}{dE}dE.\;
\label{Sradio}
\end{equation}
Here, $s$ is the line-of-sight distance and the factor 2 indicates the contributions of both high-energy electrons and positrons in the synchrotron radiation emission.

To calculate the radio contribution from dark matter annihilation, we follow two possible dark matter distribution in M33, the Navarro-Frenk-White (NFW) density profile \citep{Navarro} and the Burkert density profile \citep{Burkert} which are, respectively, expressed as
\begin{equation}
\rho_{\rm DM}(r)=\frac{\rho_{\rm s}}{\frac{r}{r_{\rm s}}\left(1+\frac{r}{r_{\rm s}}\right)^2},
\end{equation}
with the best-fit parameters, $\rho_{\rm s}=1.79^{+0.77}_{-0.61} \times 10^{-25}$ g/cm$^3$ and $r_{\rm s}=6.42^{+0.56}_{-0.47}$ kpc \citep{Fune},
and
\begin{equation}
\rho_{\rm DM}(r) =\frac{\rho_{\rm c}}{\left(1+\frac{r}{r_{\rm c}}\right)\left(1+ \frac{r^2}{r_{\rm c}^2} \right)}
\end{equation}
with $\rho_{\rm c}=12.1\pm2.0 \times 10^{-25}$ g/cm$^3$ and $r_{\rm c}=7.5\pm1.5$ kpc \citep{Corbelli}.

For a typical dark matter annihilation model, $m_{\rm DM}$ and $\langle \sigma v \rangle$ are the only free parameters. Nevertheless, standard cosmology predicts that $\langle \sigma v \rangle=2.2 \times 10^{-26}$ cm$^3$/s if dark matter particles were produced thermally at the early epoch \citep{Steigman}. This standard scenario will be particularly examined.

Apart from the dark matter contribution, the baryonic contribution is also very important for the radio flux density. We assume that the radio flux density contributed by the baryonic component is proportional to the baryonic density. As mentioned above, we will not analyze the bulge component as the data points are too few. We will only focus on the disk region. Observations show that the disk surface brightness profile can be best-fitted with an exponential function in $-r$ \citep{Seigar}. This can be transformed to the similar functional form of the line-of-sight baryonic contribution:
\begin{equation}
S_{\rm b}(r,\nu)=k(\nu)e^{-r/h},
\end{equation}
where $k(\nu)$ is a constant depending on frequency $\nu$ only, and $h=1.70 \pm 0.12$ kpc is the disk scale length \citep{Seigar}. The total radio contribution would be given by
\begin{equation}
S_{\rm total}(r,\nu)=S_{\rm DM}(r,\nu)+S_{\rm b}(r,\nu).
\end{equation}

\section{Data analysis}
Let's consider the dark matter-only model first (i.e. there is no baryonic contribution to the radio signal). For a particular frequency $\nu_j$, the goodness of fits can be determined by the reduced $\chi^2$ value:
\begin{equation}
\chi_{{\rm red},j}^2=\frac{1}{N-p}\sum_i\left[\frac{S_{\rm total}(r_i,\nu_j)-S_{\rm obs}(r_i,\nu_j)}{\sigma_{\rm obs}(r_i,\nu_j)}\right]^2,
\end{equation}
where $S_{\rm obs}(r_i,\nu_j)$ and $\sigma_{\rm obs}(r_i,\nu_j)$ are the data of the observed flux density and uncertainties of the observed flux density at different angular radii $r_i$ respectively, $N=19$ is the number of data points for each $\nu_j$, and $p$ is the number of free parameters in the fitting. We can also combine the reduced $\chi^2$ values for different frequencies to get the total $\chi^2$ value:
\begin{equation}
\chi^2=\sum_j \sum_i \left[\frac{S_{\rm total}(r_i,\nu_j)-S_{\rm obs}(r_i,\nu_j)}{\sigma_{\rm obs}(r_i,\nu_j)}\right]^2.
\end{equation}

First of all, we examine the standard cosmological scenario (i.e. $\langle \sigma v \rangle=2.2 \times 10^{-26}$ cm$^3$/s) for both dark matter density profiles. In Fig.~3, we show the $\chi_{{\rm red},j}^2$ and the total $\chi^2$ as a function of dark matter mass for $m_{\rm DM} \ge 5$ GeV. We can see that there exist some minimum $\chi^2$ values which indicate the best fits, especially for the $e$ and $\mu$ channels. However, the total $\chi^2$ values for these scenarios are larger than 700, which means that the dark matter-only model cannot explain the radio data of M33. In Fig.~4, we fit the corresponding flux profiles with the best-fit scenarios and we can see that the fits are very poor indeed for both NFW and Burkert profiles. Therefore, the dark matter-only model following the standard cosmological annihilation cross section is unlikely to account for the radio flux profiles.

Then we consider the baryon-only model (i.e. there is no dark matter contribution to the radio signal). For each frequency, we have a unique proportional constant $k$. By minimizing the reduced $\chi^2$ values, we can obtain the best-fit value of $k$ for each frequency. In fact, many radio studies usually assume that the radio flux and the frequencies have a power-law relation \citep{Tabatabaei}. This implies that the proportional constant $k$ for different frequencies can be theoretically connected with a power law in frequencies as well. However, there are some cases in which the radio flux spectrum deviates significantly from a power-law description \citep{Lisenfeld}. Therefore, we assume that the constant $k$ is a completely free parameter for each frequency to minimize the possible systematic error involved. In Fig.~4, we show the best-fit flux profiles with the best-fit values of $k$. The reduced $\chi^2$ values are 0.47, 0.11 and 0.04 for $\nu=2.70$ GHz, 4.85 GHz and 8.35 GHz respectively, which are much smaller than the reduced $\chi^2$ values for the dark matter-only model. Therefore, the baryon-only model can give much better fits compared with the dark matter-only model (see Fig.~4). In other words, the dark matter contribution may be very small compared with the baryonic contribution.

Now we test a more realistic combined model: dark matter plus baryonic contribution. For the standard cosmological scenario (i.e. $\langle \sigma v \rangle=2.2 \times 10^{-26}$ cm$^3$/s), adding dark matter component does not have a large impact on the total $\chi^2$ values (see Fig.~5). There exist some minimum values of $\chi^2$ for the $e$ channel ($m_{\rm DM}=5$ GeV with Burkert profile) and $\mu$ channel ($m_{\rm DM}=10$ GeV with Burkert profile). However, the statistical significance of these two best-fit cases is very small ($<1.3 \sigma$). Generally speaking, as the dark matter contribution is relatively small, $m_{\rm DM} \ge 5$ GeV is allowed for all annihilation channels with the thermal annihilation cross section.    

Beside assuming the thermal annihilation cross section, we can set the annihilation cross section as a free parameter and obtain the $2\sigma$ upper bounds of the annihilation cross section for different $m_{\rm DM}$ and annihilation channels. The annihilation cross section might be larger or smaller than the standard value if dark matter was not thermally produced, or the annihilation cross section is velocity-dependent \citep{Yang}. For the dark matter-only model, we can find the best-fit annihilation cross section for different $m_{\rm DM}$ by minimizing the values of $\chi_{\rm red}^2$. In Fig.~6, we show the best-fit annihilation cross section for each frequency and the case of combining three frequencies. The values of the best-fit cross section are larger than the standard thermal annihilation cross section. In the $\chi^2$ plots shown in Fig.~7, we can see that there exist two cases where the $\chi^2$ is minimum: $m_{\rm DM}=10$ GeV for the $e$ channel (NFW profile with $\langle \sigma v \rangle=1.3\times 10^{-24}$ cm$^3$/s) and $m_{\rm DM}=40$ GeV for the $\mu$ channel (Burkert profile with $\langle \sigma v \rangle=1.3 \times 10^{-23}$ cm$^3$/s). Although the total $\chi^2$ values for these two cases are slightly larger than the value in the baryon-only model, they represent very good fits for the data. We show the corresponding fits of the radio data in Fig.~8. Nevertheless, when comparing the best-fit results for this dark matter-only model with the AMS-02 positron and gamma-ray analyses, they are excluded by the $2\sigma$ upper limit bands on $\langle \sigma v \rangle$ \citep{Ackermann,Cavasonza}. For instance, the $2\sigma$ upper limits derived from the gamma-ray data of the Milky Way dwarf galaxies are $\sim 2 \times 10^{-26}$ cm$^3$/s ($m_{\rm DM}=10$ GeV) and $\sim 9\times 10^{-25}$ cm$^3$/s ($m_{\rm DM}=40$ GeV) respectively for the $e$ channel and $\mu$ channel \citep{Ackermann}. Note that the unknown functional form of the dark matter density assumed in the gamma-ray analysis (i.e. the so-called J-factor) and whether the annihilation cross section is velocity-dependent would contribute large systematic uncertainties in the derived bounds \citep{Chiappo,Boucher}. 

Now we consider the combined model (dark matter plus baryons) again. In principle, we can also obtain the best-fit annihilation cross section by tuning the value of $k(\nu)$ for each frequency. However, as the baryon-only model can provide excellent fits to the radio data, adding a dark matter component does not improve the fits significantly. There exist multiple degenerate best-fit annihilation cross sections for different values of $m_{\rm DM}$. This suggests that the contribution of dark matter annihilation in the disk region must be small compared with the baryon emission. Nevertheless, we can obtain the $2\sigma$ upper bounds of the annihilation cross section for different $m_{\rm DM}$ and annihilation channels. By adopting the baryon-only model described above, we investigate on how large of the dark matter contribution would exceed the $2\sigma$ margins. In Fig.~9, following the combined model, we show the overall $2\sigma$ upper limits of the annihilation cross section against $m_{\rm DM}$ for different channels by combining the radio data of the three frequencies. These are the strongest limits for dark matter annihilation cross section in the disk region of M33 galaxy as we have included baryon contribution in the radio emission. 

\section{Discussion}
In this article, we present an analysis combining the radio flux profile and multi-wavelength approach to get more comprehensive constraints on dark matter annihilation. Here, we use the archival data of the M33 galaxy to get the radio flux profiles for $\nu=2.70$ GHz, 4.85 GHz and 8.35 GHz. We find that the baryon-only model can give much better fits compared with the dark matter-only model. This means that the overall radio emission in the M33 disk is likely dominated by baryon-related emission (i.e. thermal and non-thermal emissions), but not dark matter contribution. Nevertheless, if annihilating dark matter dominates the radio emission, we find that dark matter with $m_{\rm DM}=10$ GeV annihilating via the $e$ channel and $m_{\rm DM}=40$ GeV annihilating via the $\mu$ channel can give the best fits for the radio data. However, the best-fit annihilation cross sections are excluded by the $2\sigma$ upper limit bands on $\langle \sigma v \rangle$ \citep{Ackermann,Cavasonza}.

Furthermore, we apply the combined model to include both contributions of dark matter and baryons. Although there is no best-fit scenario for dark matter due to the degeneracy problem, we can constrain the $2\sigma$ conservative upper bounds of the annihilating cross section for different annihilation channels. Compared with the ANTARES neutrino bound \citep{Albert}, our upper limits are tighter for the $b$ and $W$ channels, and the $\mu$ channels with $m_{\rm DM} \le 100$ GeV. Also, our limit for the $\mu$ channel with $m_{\rm DM} \sim 100$ GeV is generally more stringent than that in the AMS-02 antiproton analysis \citep{Calore2}. However, our limits for the $b$, $\tau$ and $W$ channels are less stringent than that in gamma-ray analysis \citep{Abdallah,McDaniel}. Overall speaking, analyzing the radio flux profile data with different frequencies can give good limits for leptophilic channels, especially for the $\mu$ channel. The results for the $\mu$ channel are particularly important as recent studies have paid more attention to the $\mu$-related annihilation \citep{Ghosh,Abdughani} due to the muon $(g-2)$ anomaly in the measurement \citep{Abi}. For the $b$ and $W$ channels, gamma-ray detection can generally provide more stringent constraints. Note that for all analyses of dark matter annihilation cross section upper limits, there are different assumptions and uncertainties involved such as dark matter density profile, cosmic-ray propagation parameters, magnetic fields, etc. Therefore, there is no single model-independent robust upper bound for the annihilation cross section. Nevertheless, all of the bounds can be complementary to each other to give more comprehensive constraints for dark matter annihilation. Our analysis using radio data of M33 can provide one of the important contributions. Moreover, compared with the previous studies of the M33 galaxy using a single frequency radio data \citep{Chan6}, the constraints in this study are more realistic and authentic for annihilating dark matter.

Note that we did not consider any boost factor due to dark matter substructures in our analysis \citep{Moline,Chan6}. Besides, our $2\sigma$ upper limits of the annihilation cross section are based on the combined model, but not the dark matter-only model. Therefore, in principle, our constraints presented here are stronger compared with that without considering baryonic contributions, provided that the baryonic model assumed is physical. For the case of assuming the annihilation cross section as a free parameter, our results are useful to constrain the non-thermal annihilating dark matter and the velocity-dependent annihilation cross section. For instance, the radio $2\sigma$ upper limits of the annihilation cross section shown in our study are generally weaker than that in \citet{Beck}. However, the constraints in current study are derived from a small galaxy while the constraints in \citet{Beck} are derived from galaxy clusters. If the annihilation cross section is velocity-dependent, such as Sommerfeld enhanced \citep{Sommerfeld,Yang}, the constraints derived in small galaxies and galaxy clusters would have different implications because the dark matter velocity dispersion in a small galaxy is much smaller than that in a galaxy cluster. 

In this study, we demonstrate a comprehensive approach in analysing the radio flux profiles with different frequencies. The overall uncertainties of the radio flux density observed for M33 galaxy are quite significant so that we cannot get very stringent constraints for dark matter. Since there is a resolution limit (i.e. the minimum beam size) for radio observations, we can only obtain good quality radio flux profile data (with small uncertainties) for nearby galaxies (e.g. Local group galaxies). If one can obtain good quality radio map with different frequencies for nearby galaxies, we can follow this comprehensive analysis to get more stringent constraints for annihilating dark matter.

\begin{figure}
\begin{center}
\vskip 3mm
\includegraphics[width=140mm]{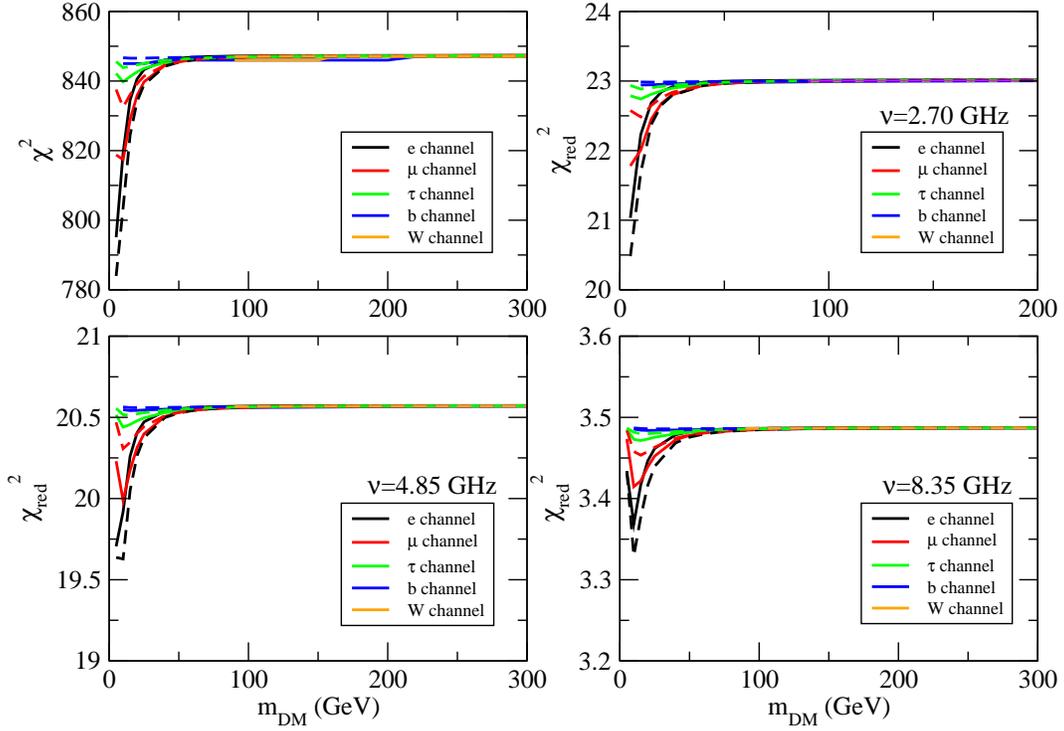}
\caption{The graph at the left upper corner indicates the total $\chi^2$ as a function of $m_{\rm DM}$ for different annihilation channels in the dark matter-only model with the thermal annihilation cross section. The other three graphs are the reduced $\chi^2$ for different frequencies. The colored solid and dashed lines indicate the dark matter density distributions following the NFW and Burkert profiles respectively.}
\label{Fig3}
\end{center}
\end{figure}

\begin{figure}
\begin{center}
\vskip 3mm
\includegraphics[width=140mm]{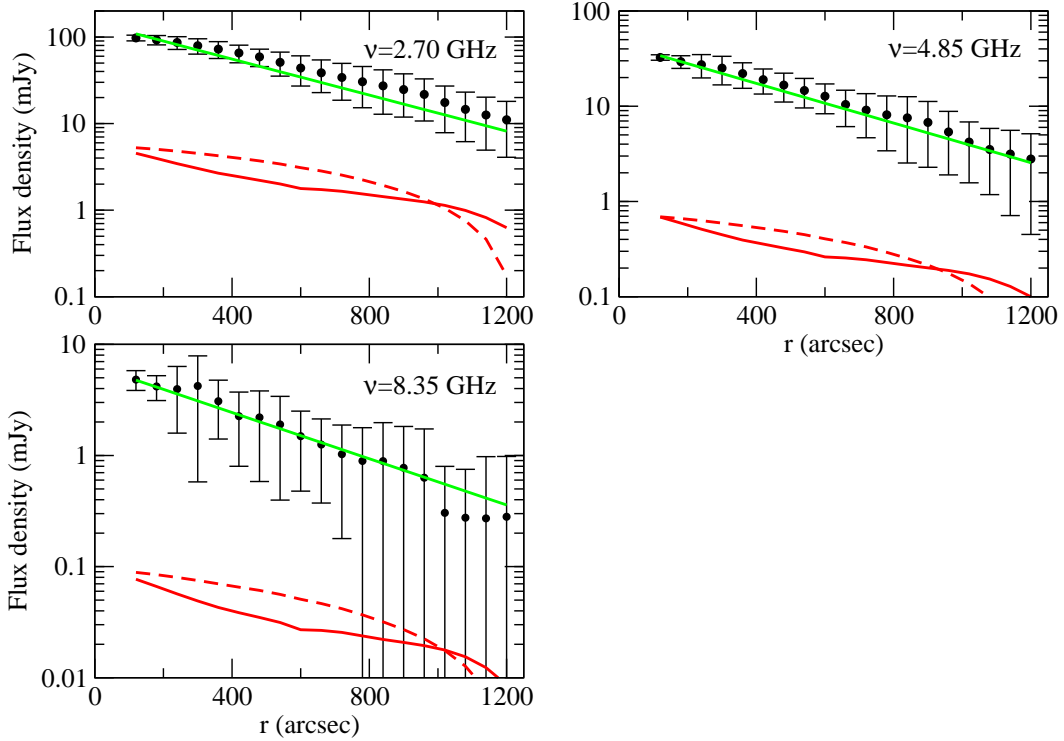}
\caption{The best-fit radio flux density profiles for the dark matter-only model (red) and baryon-only model (green). The red solid and dashed lines indicate the dark matter density distributions following the NFW density profile ($\nu=2.70$ GHz and $\nu=4.85$ GHz: $m_{\rm DM}=5$ GeV, $\nu=8.35$ GHz: $m_{\rm DM}=10$ GeV) and Burkert density profile ($\nu=2.70$ GHz: $m_{\rm DM}=5$ GeV, $\nu=4.85$ GHz and $\nu=8.35$ GHz: $m_{\rm DM}=10$ GeV) respectively. For the dark matter-only model, the thermal annihilation cross section is assumed and all of the best-fit scenarios are via the $e$ channel. Here, the data points for the bulge region are not included.}
\label{Fig4}
\end{center}
\end{figure}

\begin{figure}
\begin{center}
\vskip 3mm
\includegraphics[width=140mm]{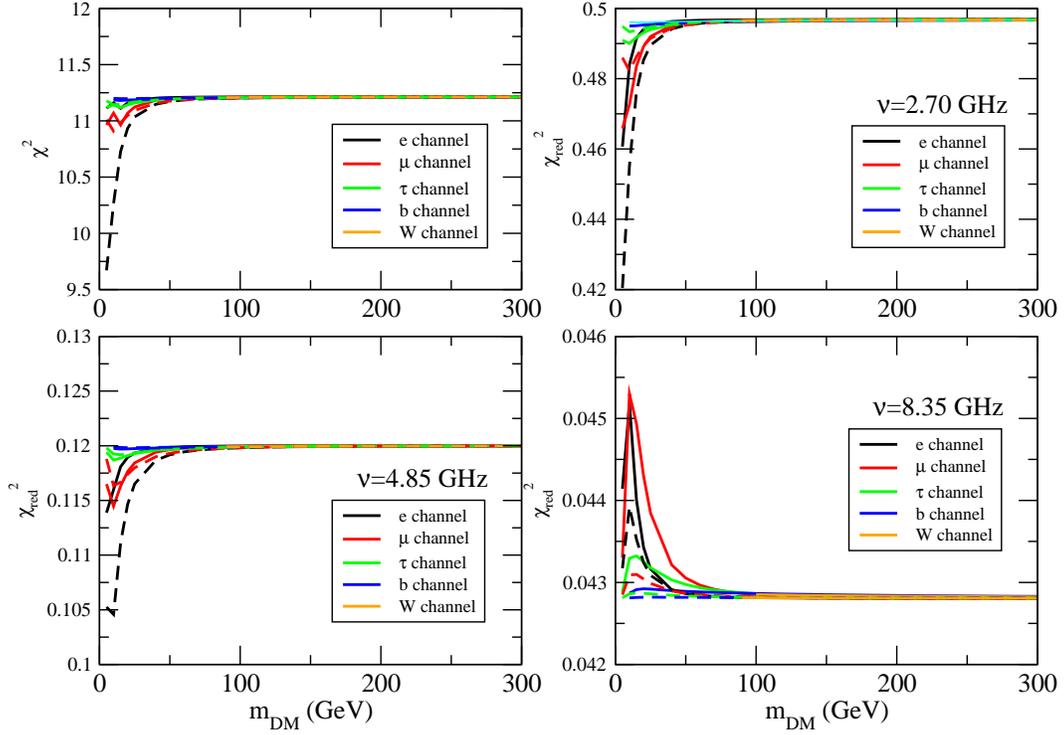}
\caption{The graph at the left upper corner indicates the total $\chi^2$ as a function of $m_{\rm DM}$ for different annihilation channels in the combined model with the thermal annihilation cross section. The other three graphs are the reduced $\chi^2$ for different frequencies. The colored solid and dashed lines indicate the dark matter density distributions following the NFW and Burkert profiles respectively.}
\label{Fig5}
\end{center}
\end{figure}

\begin{figure}
\begin{center}
\vskip 3mm
\includegraphics[width=140mm]{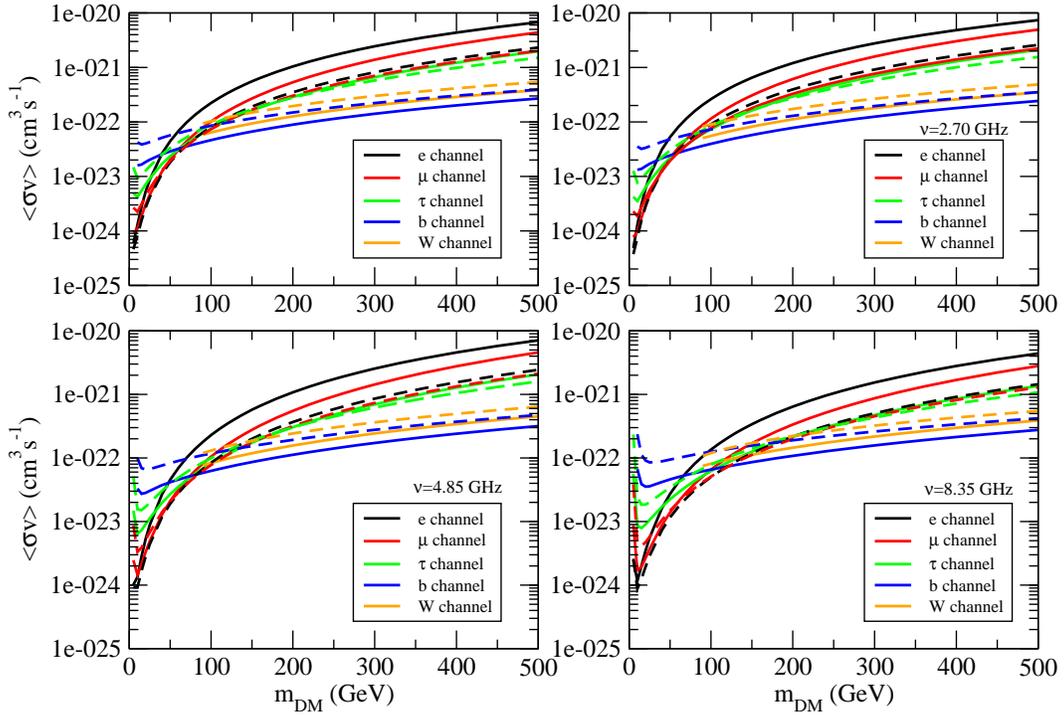}
\caption{The graph at the left upper corner indicates the best-fit annihilation cross sections as a function of $m_{\rm DM}$ for different annihilation channels in the dark matter-only model (considering all of the three frequencies). The other three graphs are the best-fit annihilation cross sections for different frequencies. The colored solid and dashed lines indicate the dark matter density distributions following the NFW and Burkert profiles respectively.}
\label{Fig6}
\end{center}
\end{figure}

\begin{figure}
\begin{center}
\vskip 3mm
\includegraphics[width=140mm]{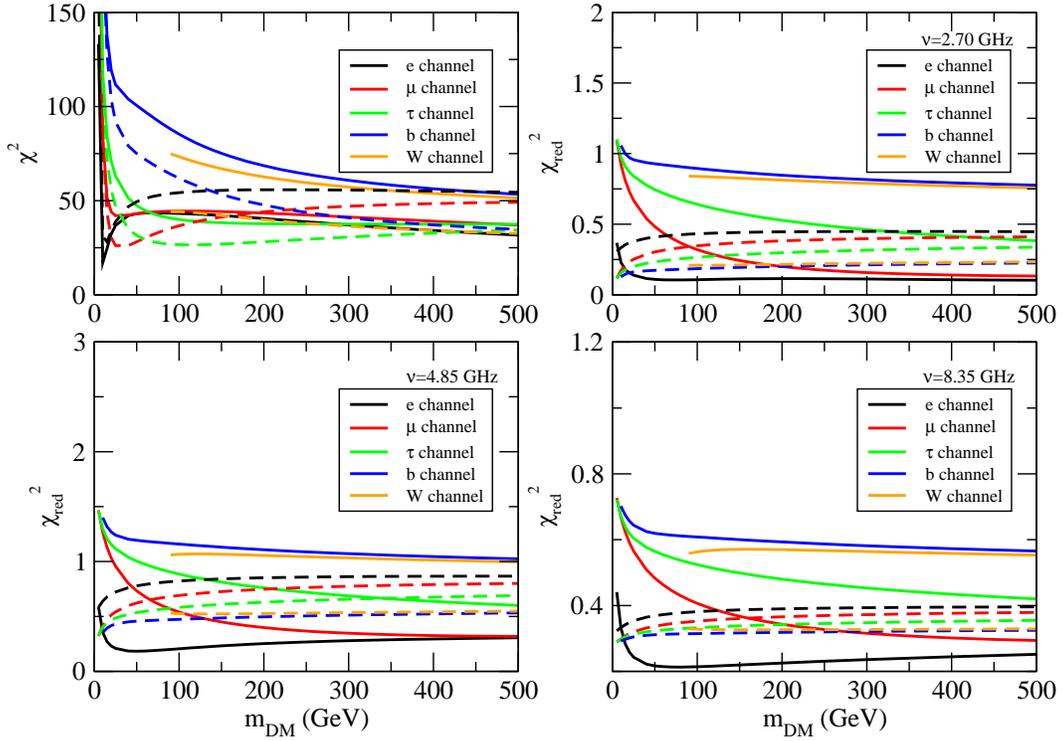}
\caption{The graph at the left upper corner indicates the total $\chi^2$ as a function of $m_{\rm DM}$ for different annihilation channels in the dark matter-only model (the annihilation cross section is a free parameter). The other three graphs are the reduced $\chi^2$ for different frequencies. The colored solid and dashed lines indicate the dark matter density distributions following the NFW and Burkert profiles respectively.}
\label{Fig7}
\end{center}
\end{figure}

\begin{figure}
\begin{center}
\vskip 3mm
\includegraphics[width=140mm]{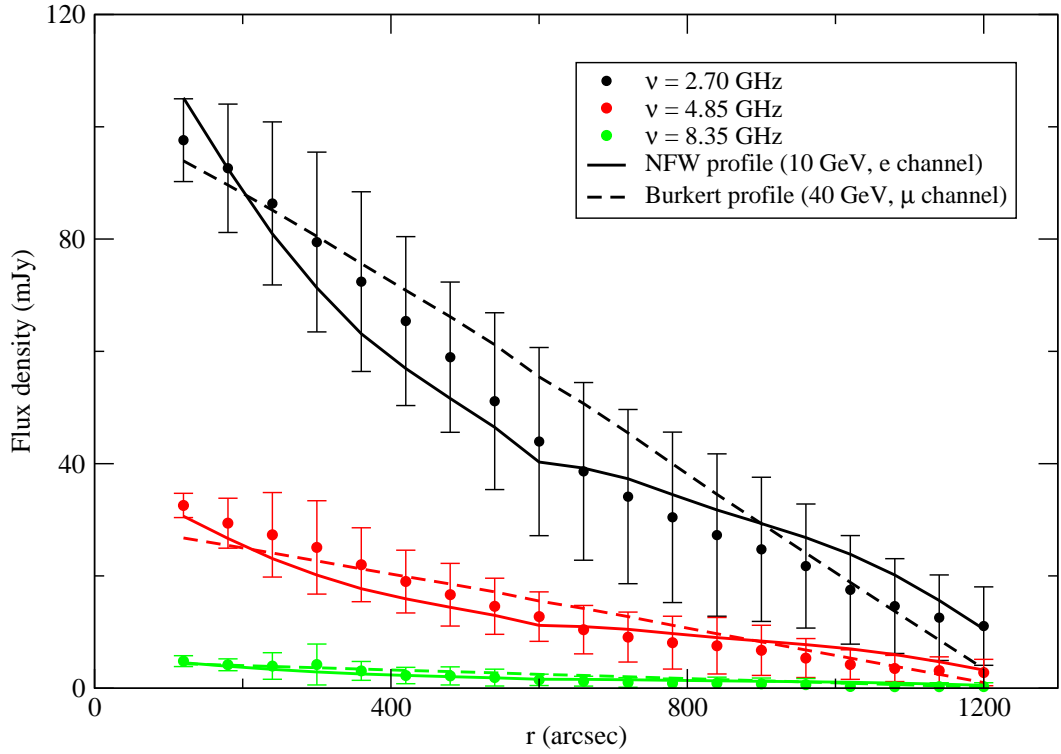}
\caption{The best-fit radio flux density profiles for the dark matter-only model (the annihilation cross section is a free parameter). The solid and dashed lines indicate the best-fit cases for the NFW profile and the Burkert profile respectively. Here, the data points for the bulge region are not included.}
\label{Fig8}
\end{center}
\end{figure}

\begin{figure}
\begin{center}
\vskip 3mm
\includegraphics[width=140mm]{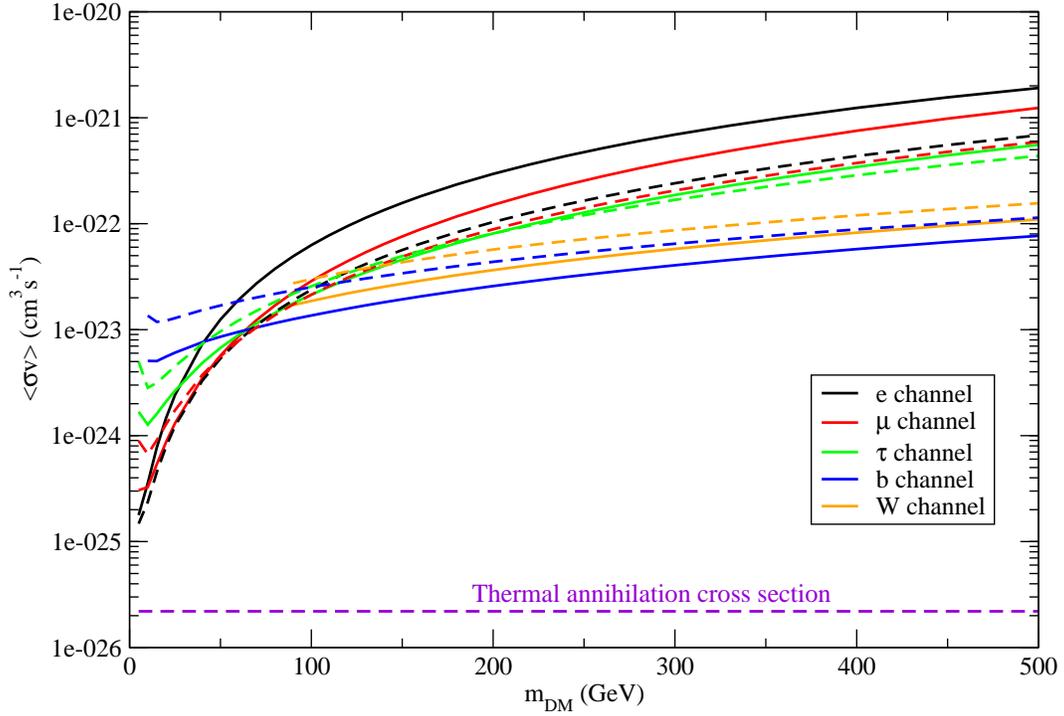}
\caption{The $2\sigma$ upper bounds of the annihilation cross section for different annihilation channels in the combined model. The colored solid and dashed lines indicate the dark matter density distributions following the NFW and Burkert profiles respectively.}
\label{Fig9}
\end{center}
\end{figure}

\section{Acknowledgements}
The work described in this paper was partially supported by a grant from the Research Grants Council of the Hong Kong Special Administrative Region, China (Project No. EdUHK 18300922). Lang Cui is thankful for the grants supported by the National SKA Program of China (No. 2022SKA0120102) and the CAS `Light of West China' Program (No. 2021-XBQNXZ-005).

\end{document}